\documentstyle[11pt,aaspp,psfig]{article}

\begin{document}

\title{\large\bf
 Luminosity Functions of Elliptical Galaxies at $z < 1.2$}
\author{\bf Myungshin Im\altaffilmark{1}, Richard E. Griffiths\altaffilmark{1},
 Kavan U. Ratnatunga\altaffilmark{1}, \\
 AND Vicki L. Sarajedini\altaffilmark{2}}
\altaffiltext{1}{Department of Physics \& Astronomy, Johns Hopkins University,
 Baltimore, MD 21218; myung, griffith, kavan@pha.jhu.edu}
\altaffiltext{2}{Steward Observatory, University of Arizona, Tucson, AZ 85721;
 vicki@as.arizono.edu}


\begin{abstract}

The luminosity functions of E/S0 galaxies are constructed in 3 different
redshift bins ($0.2 < z < 0.55$, $0.55 < z < 0.8$, $0.8 < z < 1.2$),
using the data from the Hubble Space Telescope Medium Deep Survey (HST
MDS) and other HST surveys.  These independent luminosity functions show
the brightening in the luminosity of E/S0s by about $0.5 \sim 1$
magnitude at $z \sim 1$, and no sign of significant number evolution.
 This is the first direct measurement of the luminosity evolution of E/S0
galaxies, and our results support the hypothesis of a high redshift of
formation ($z > 1$) for elliptical galaxies, together with weak
evolution of the major merger rate at $z < 1$.

\end{abstract}


{\it Submitted to ApJ.Lett., Dec. 23 1995, Accepted Feb. 5 1996}

\section{Introduction}

The images of faint galaxies taken with the Wide Field and Planetary
Camera (WFPC2) on the Hubble Space Telescope (HST) have provided
invaluable information on the morphology of galaxies when the universe
was about half its present age.  With these data, reliable morphological
classification is possible down to a magnitude limit of $I < 22 \sim
23$, and we have identified the morphological nature of the galaxies
which dominate the number counts at faint magnitudes (Im et al. 1995a,
1995b; Griffiths et al. 1994; Casertano et al. 1995;
Driver, Windhorst, \& Griffiths 1995; Glazebrook et al. 1995;
Abraham et al. 1996).
  For E/S0 galaxies, Im et al. (1995c)
find evidence for mild luminosity evolution and no strong merging
activity at $z < 1$, using the observed size and colour distributions.
Likewise, the HST observations of E/S0 galaxies in clusters
at moderate and high redshifts show little sign of strong luminosity
evolution (Dickinson 1995).

Lilly et al. (1995a) and Ellis et al. (1995) have shown luminosity
functions of the total or colour-divided galaxy population at different
redshifts and tried to constrain the evolution of faint galaxies up to
$z~\sim~1$.  In particular, Lilly et al. (1995a) divided the galaxies
into two subsamples according to their colors (red or blue), and then
showed that the luminosity of red galaxies evolves mildly, in contrast
with the blue galaxies.  This result on the red galaxies is consistent
with the HST observations of ellipticals, but because of the lack of
morphological classification for most of the galaxies in the Lilly et
al. sample, detailed constraints on the luminosity evolution of E/S0
galaxies have not been obtained in their study.  The HST data can
provide such morphological information, and we will thus construct
luminosity functions of {\it E/S0 galaxies} at moderate to high redshifts
in order to study their evolution.

\section{Data}

Our selection of elliptical galaxies comes from the full HST MDS which
observes random fields using WFPC2, and we have also included
generically similar archival data from a strip of sky originally
surveyed in primary mode by Groth et al. (1994).  All of these fields
were observed in both the $V$ and $I$ bands (F606W and F814W).  The
detection limit for each field covers the range $I\simeq 23 \sim 25$ and
$V\simeq 24 \sim 25.5$.  For each object detected, the observed image is
fitted with simple model profiles (point source, $r^{1/4}$ profile and
exponential profile) using a 2-dimensional maximum likelihood technique
(Ratnatunga et al 1995).  We identify E/S0 galaxies via (i) morphology
and (ii) luminosity profile. Point (ii) is an important step which is
needed in order to exclude dwarf elliptical galaxies (dE) from our
sample. The dE galaxies have different photometric properties from
normal ellipticals (e.g, see Im et al. 1995b and references therein).
 From 56 HST WFPC2 fields, we find 376 elliptical galaxies at $18 < I <
22$, the same sample as that used in the previous study of number counts
and constraints on cosmological parameters using these galaxies (Im et
al. 1995c).  For a dozen ellipticals in our sample, spectroscopic
observations were obtained with the KPNO 4 meter using the Cryogenic
Camera and RC Spectrograph with multi-slit masks during observing runs
in October 1994 and April 1995.  Sufficiently high signal-to-noise
spectra were obtained during 2 to 3 hour exposures in order
to measure redshifts.
 Redshifts were generally determined through identification of
 the CaII H\&K absorption
feature in each spectrum, at a rest wavelength of $\sim 4000\AA$.
Redshifts for 11 E/S0s in the archived HST data from the Groth-Westphal
field at 14H+52$^o$ were taken from Lilly et al.(1995b).
 These latter redshifts were extremely useful since they covered the range
 from $z \sim 0.6$ to $z \sim 1.2$.

\begin{figure}[bth]
\centerline{\psfig{figure=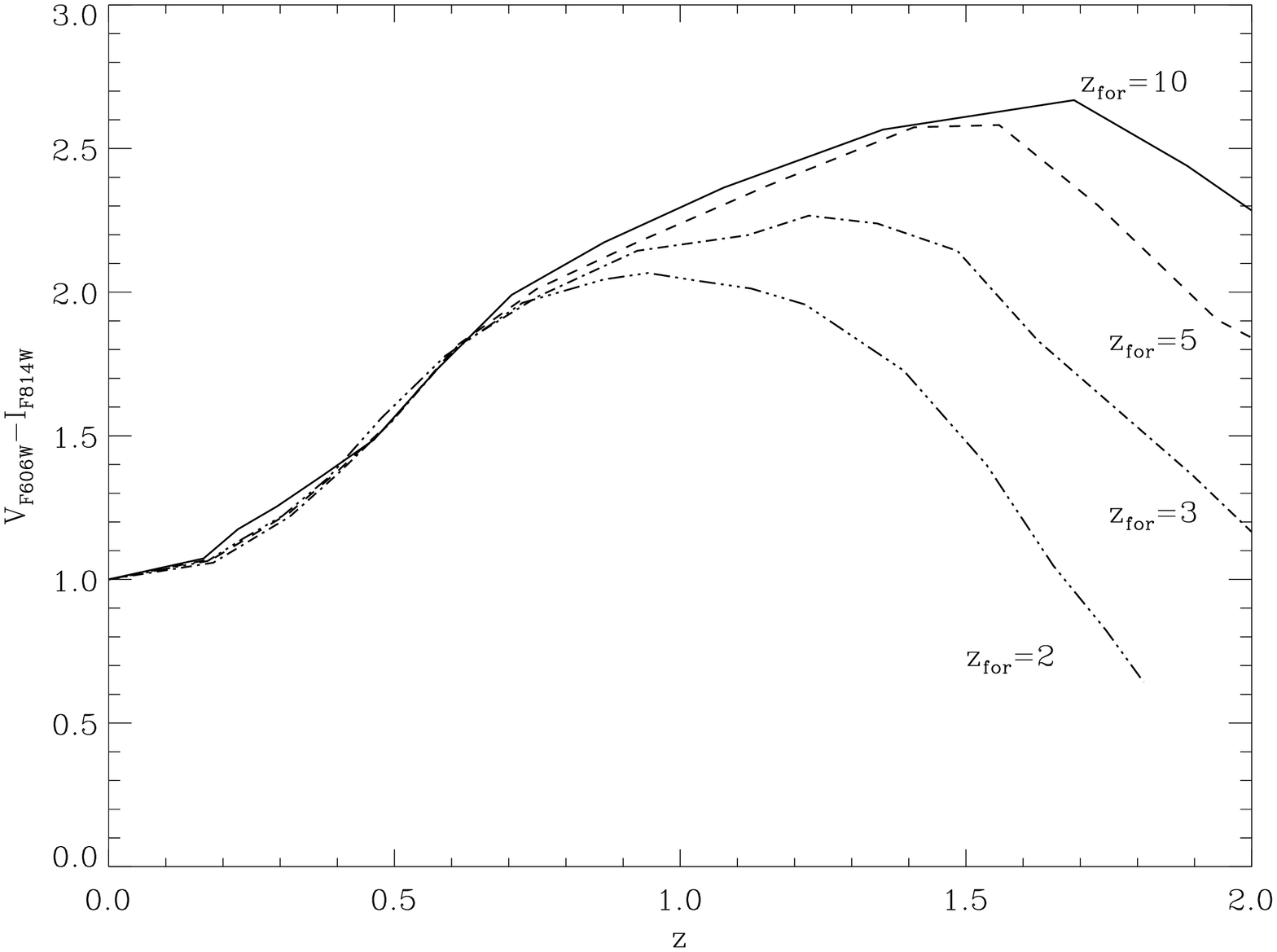,width=4.in,angle=90}}
\footnotesize{
Fig. 1: The predicted $V-I$ colors of E/S0 galaxies vs. redshift.  Note
that the $z - (V-I)$ relation at $z < 1$ is nearly independent of
the formation redshift $z_{for}$.}
\end{figure}

\section{Estimate of Photometric Redshifts}

\begin{figure}[htb]
\centerline{\psfig{figure=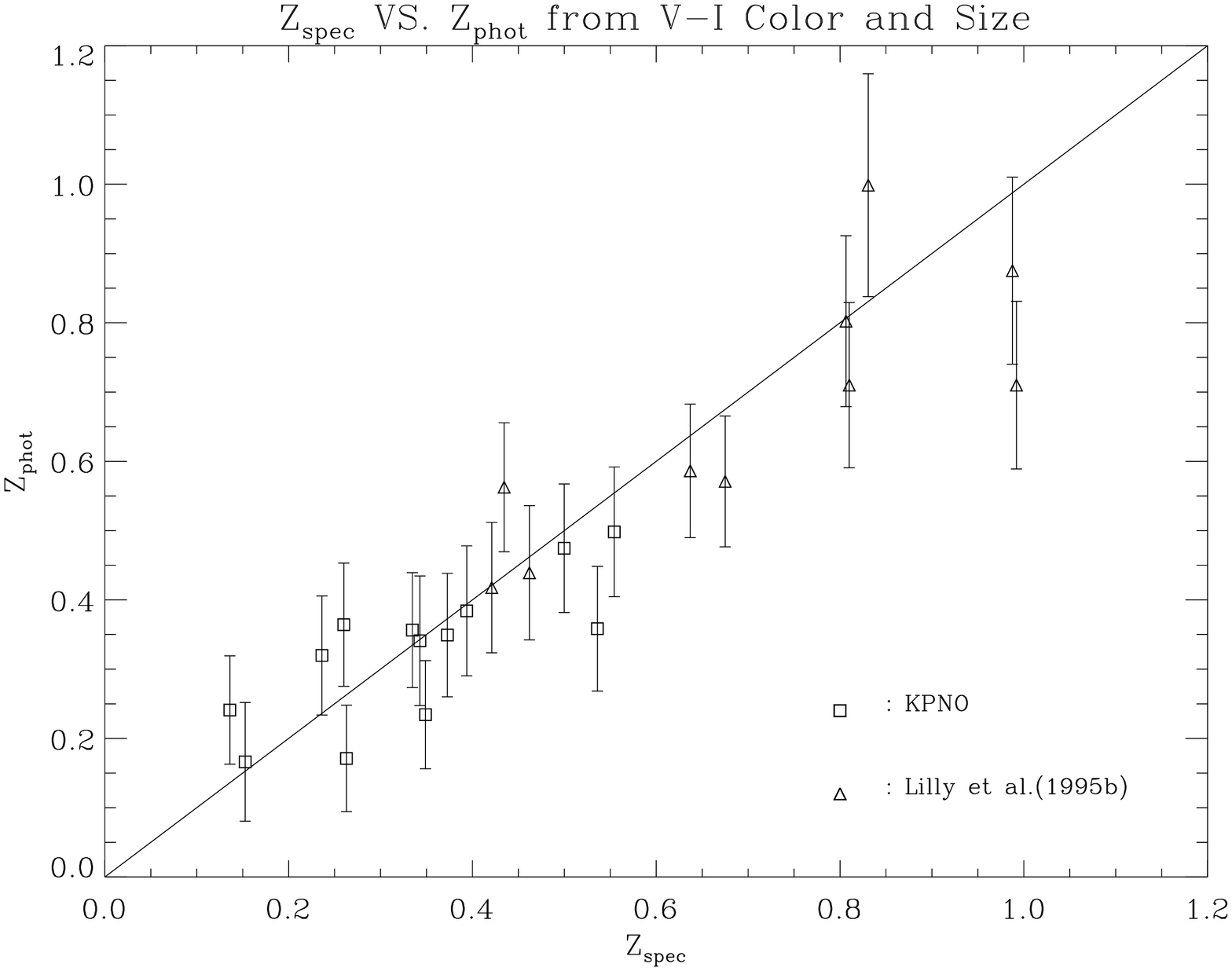,width=4.in,angle=90}}
\footnotesize{
 Fig. 2: The measured spectroscopic redshifts ($z_{spec}$) vs.
($z_{phot}$), redshifts estimated photometrically using the $V-I$
colors, the apparent magnitude and the half light radii of E/S0
galaxies.  The accuracy of $z_{phot}$s is about $\delta z_{phot} < 0.1$
at $z < 0.5$ and $\delta z_{phot} \sim 0.13$ at $z \sim 1$. }
\end{figure}

Previous studies have shown that the spectral energy distribution of
present-day normal E/S0 galaxies can be well fitted by a single
starburst model with an age of more than 10 Gyr (Bruzual \& Charlot 1993;
 Arimoto \& Yoshii 1986; Guiderdoni \& Rocca-Volmerange 1987).  According to
these models, the color of E/S0 galaxies is quite insensitive to
luminosity evolution up to $z \sim 1$ (Fig.~1).  We therefore use the
$V-I$ color of E/S0 galaxies to estimate redshifts. The use of the HST
$V$ and $I$ bands is advantageous in this respect since they are
relatively insensitive to the detailed shape of individual spectra,
owing to their broad widths.  We also use the half light radii of the
E/S0s in order to complement the photometric estimation of redshifts.
For a given apparent magnitude, galaxies with lower surface brightness
(or larger angular sizes) are likely to be at higher redshifts than
galaxies with higher surface brightness (or smaller angular sizes)
because of the $(1+z)^{4}$ surface brightness dimming effect (Im et
al. 1995a).

In Fig.2, we show estimated photometric redshifts vs. spectroscopic
redshifts for a subsample of E/S0 galaxies.
 The photometric redshifts estimated
using colors and sizes are accurate to within an error of about 0.08
at low redshifts and about 0.13 at high redshifts.

In order to calculate absolute magnitudes, the K-correction has been
applied, but no evolutionary correction was used since our intention is
to measure the degree of evolution of the luminosity
function.  The error in $z_{phot}$ leads to an uncertainty of about 0.5
mag in the absolute magnitudes at $0.2 < z < 1.2$.

\section{Luminosity function of elliptical galaxies at $z < 1.2$}

\begin{figure}[htb]
\centerline{\psfig{figure=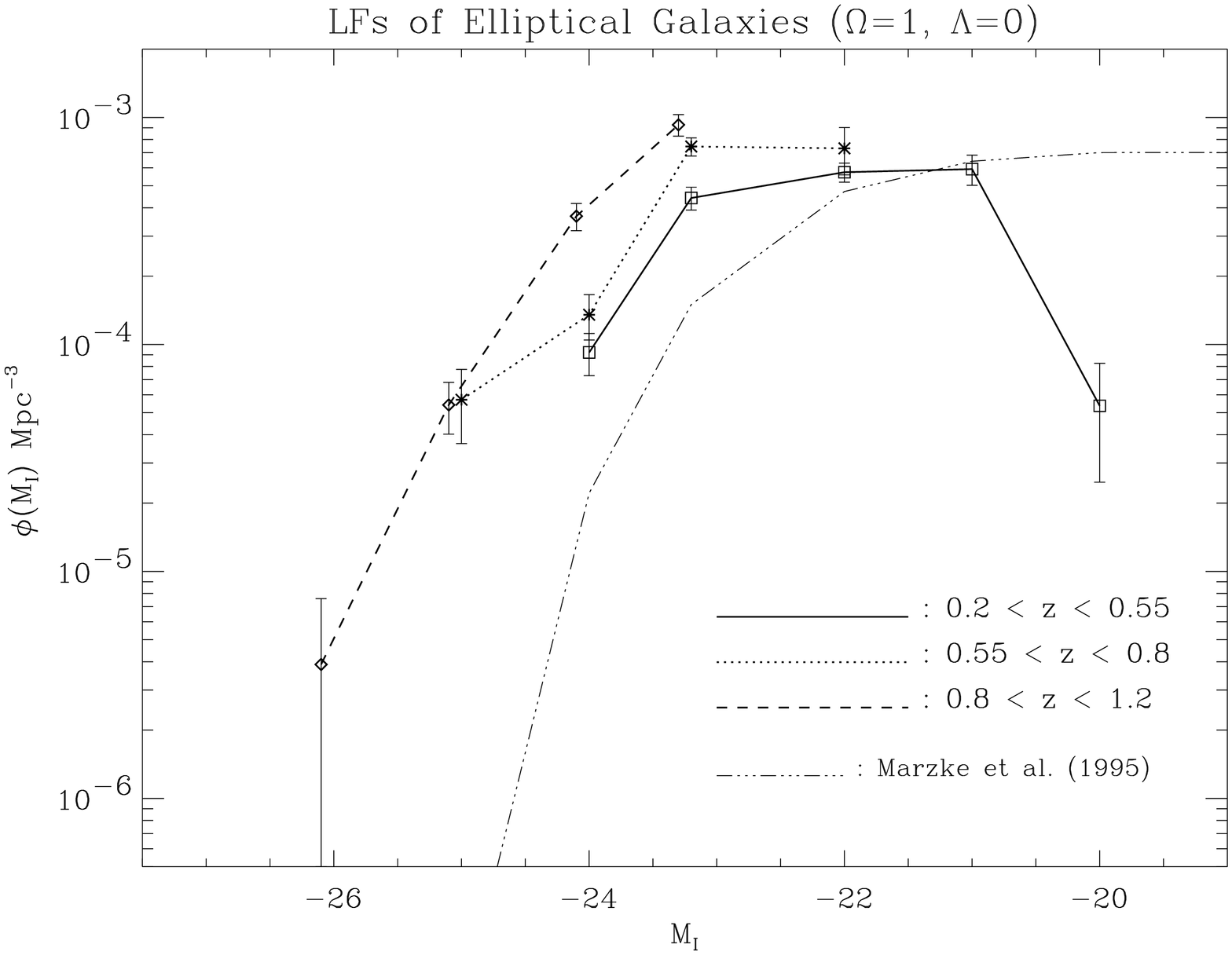,width=4.in,angle=90}}
\footnotesize{
Fig. 3: The luminosity function of E/S0 galaxies at i) $0.2 < z < 0.55$
(solid line), ii) $0.55 < z < 0.8$ (dotted line), and iii) $0.8 < z < 1.2$
(dashed line). Cosmological parameters $\Omega_{m}=1$ and $H_{0}=50
km\,sec^{-1}\,Mpc^{-1}$ are used to derive the luminosity functions. The
luminosity function of E/S0 at $z \sim 0$ from Marzke et al. (1995) is
also plotted assuming $Z-I=2.4$. }
\end{figure}

 We have constructed the luminosity function of E/S0 galaxies using the
$V/V_{max}$ technique (Huchra \& Sargent 1973; Lilly et al.  1995a)
within three redshift bins, i) $0.2 < z < 0.55$, ii) $0.55 < 0.8$, and
iii) $0.8 < z < 1.2$ under the assumption of two different sets of
cosmological parameters: A) $\Omega_{matter} (hereafter, \Omega_{m})=1,
 ~\Lambda=0$, and B) $\Omega_{m}=0.2,~\Lambda=0.8$.
  Each redshift bin contains about 120
galaxies, and this gives about $20 \sim 30$ galaxies available within
each absolute magnitude bin for the determination of the luminosity
functions (LFs).  The calculated LFs are plotted in Fig.3. and Fig.4.,
 along with the E/S0 LF from the CfA redshift survey (Marzke et al. 1995).
  The error on each point is obtained by a bootstrap algorithm.  There is a
clear trend that the LF moves towards brighter magnitudes with
increasing redshift, i.e. we observe luminosity evolution.  Taking into
account the uncertainty of about 0.5 in the estimate of absolute
magnitudes, there thus appears to be luminosity evolution by $ 0.5 \pm
0.5$ in I magnitude from $z \sim 0.4$ to $z \sim 0.7$, and $1 \pm 0.5$
magnitude brightening from $z \sim 0.4$ to $z \sim 1$.  The
corresponding expected luminosity evolution at $z \sim 1$
 from spectral evolution
 models is about 0.7 to 1.4 ($z_{for} > 3$), consistent with our findings
 (Bruzual \& Charlot 1993; Colin et al 1993; Im et al. 1995c; Roche et al.
 1995).  It is not clear whether there is luminosity evolution from z=0
 to z=0.4, because of the uncertainty in the value of $M_{*}$
  for the local elliptical LF
(Marzke et al. 1995; Loveday et al. 1992; Zucca et al. 1994) as well as
the 0.5 magnitude uncertainty in our LF, but the models predict about
0.3 magnitude of brightening at this redshift.
   There is observational evidence for the existence of intermediate
 age stellar populations in cluster E/S0s, implying that there were secondary
 bursts of star formation (e.g., Couch \& Sharples 1987).
  Starbursting E/S0s may look more irregular and bluer than the normal
 ellipticals, which are not found in our sample.
   Hence, our result does not necessarily exclude such models of E/S0s with
 the secondary starbursts (Charlot \& Silk 1994).

\begin{figure}[htb]
\centerline{\psfig{figure=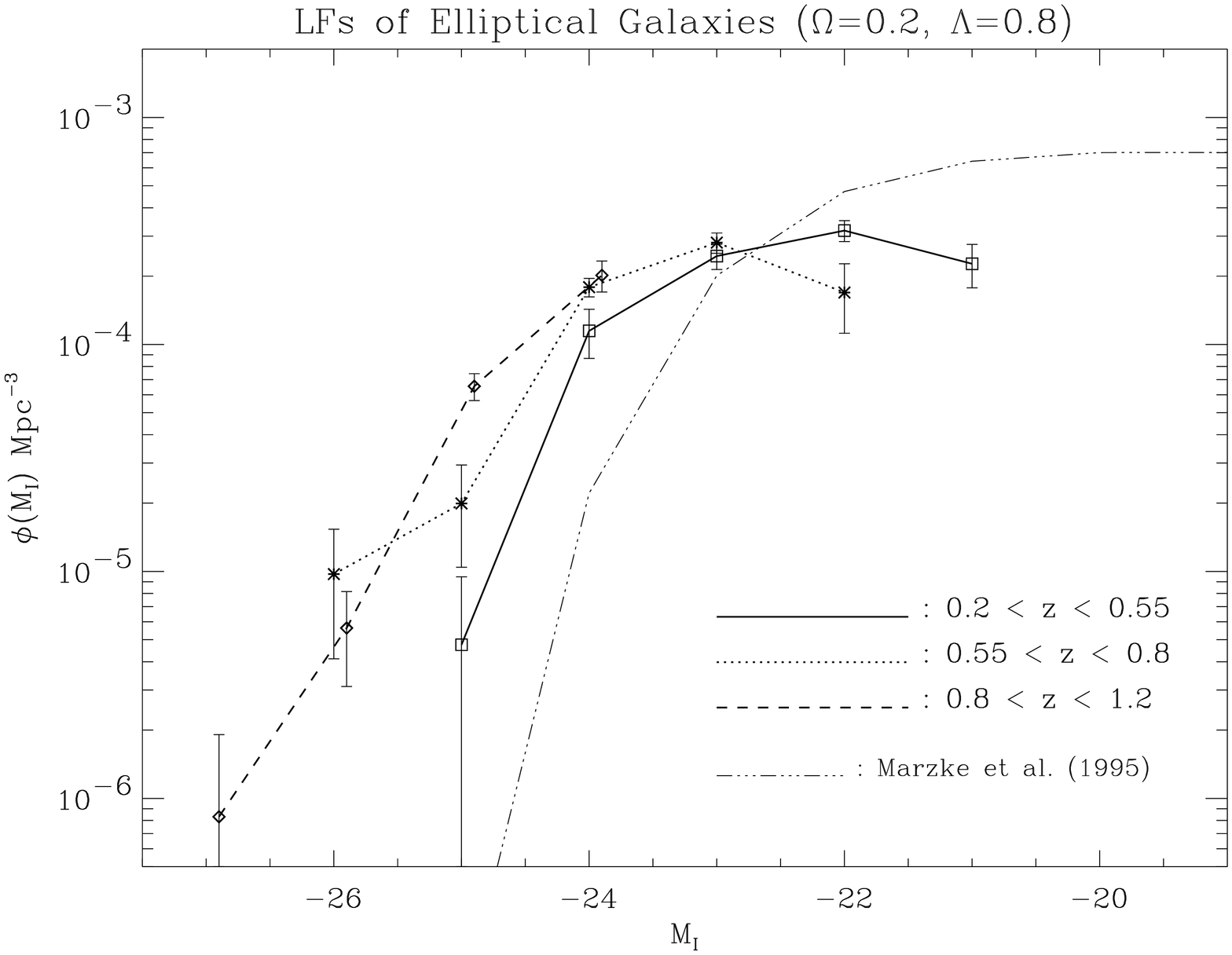,width=4.in,angle=90}}
\footnotesize{
Fig. 4: The luminosity function of E/S0 galaxies at i) $0.2 < z < 0.55$
(solid line), ii) $0.55 < z < 0.8$ (dotted line), and iii) $0.8 < z < 1.2$
(dashed line), on the assumption of $\Omega_{m}=0.2$, $\Lambda=0.8$ and
$H_{0}=50 km\,sec^{-1}\,Mpc^{-1}$. The luminosity function of E/S0 at
$z \sim 0$ from Marzke et al. (1995) is also plotted assuming $Z-I=2.4$.}
\end{figure}

  To show the completeness of our sample, we list $V/V_{max}$ values and
the number of ellipticals used in Table 1.  When $V/V_{max} = 0.5$, the
sample can be considered complete and there is then no strong bias in
the estimate of the LF (Huchra \& Sargent 1973; Zucca et al. 1994).
 In Table 1, we find that $V/V_{max} \simeq 0.5$ for most cases
 except for a few outliers, and we conclude that our result is not
 affected by any bias due to the incompleteness of the sample.
   Also, we find that the $V/V_{max}$ is distributed uniformly
 from 0 to 1 (not shown in this paper).
   The elliptical galaxies live preferentially
 in clusters and the derived LF could be biased by ellipticals in a
 few large clusters. If this is the case, the $V/V_{max}$ would be clumped
 around the certain value where the cluster lies. We do not find this kind of
 trend, thus our result is not affected by large clusters.

  In order to provide a more quantitative result, we have also estimated
parameters for the LFs using the STY method (Sandage, Tamman, \& Yahil 1979;
 Loveday et al. 1992; Marzke et al. 1995), on the assumption that
 they are described by the Schechter form (Schechter 1976).
   Because of the lack of a sufficient
number of galaxies to fit the LF over a reasonable magnitude range, the
estimated parameters ($\alpha~and~M_{*}$) are rather unstable, and the
apparent steepening of the estimated slope at high redshift should not be
considered seriously (first and second rows of Table 2).  In order to
obtain a more stable fit, we therefore used a fixed $\alpha=-0.85$
 (Marzke et al. 1995) and
estimated the change in $M_{*}$. The third row of Table 1 shows $M_{*}$
estimates by this method, and we clearly see the brightening of $M_{*}$
with redshift.
  We also note that the observed I magnitude is roughly equivalent to the
 rest frame B-magnitude at $z \sim 1$ within a few tenths of magnitude
 (Lilly et al. 1995a).
  For this reason, we constructed the luminosity function at $z \sim 1$
 without applying K-correction, and estimated the $M_{B*}$ using
$\alpha=-0.85$.
  This way, we get $M_{B*}=-20.4 \sim -21.1 + 5\,log(h)$ (note that $M_{B*}
 \simeq -19.5 +5\,log(h)$ at z=0: Efstathiou et al. 1988; Marzke et al.1995)
 consistent with our finding of the luminosity evolution in I magnitude
 (fifth row of the Table 2).

    If, as a result of mergers, the average mass of ellipticals
 is a decreasing function of look-back time, our result is also consistent
 with strong luminosity evolution ($>> 1 mag$)
 at z=1. However, this is unlikely
  since the strong merging activity accompanies
 the remarkable decrease of the number density as a function of redshift,
 which is not observed in our data (however, see discussions below
 for a possible loophole in this argument).

  In order to quantify the possible number evolution of E/S0s, we tried to
 estimate $\phi_{*}$ using the parameters in the third row of Table 2 and
 the observed LFs in Fig.3.
   The observed LFs show $\phi_{*}$ may have decreased by about
 50 \% since z=1 if $\Omega_{m}=1~and~\Lambda=0$.
   If $\Omega_{m}=0.2~and~\Lambda=0.8$, the $\phi_{*}$ value hardly changes
 as a function of redshift, probably by less than 30 \% (see the fourth row of
 Table 2).
   If the value of $\phi_{*}$ is indicative of the number density evolution
 (i.e, $M_{*}$ is fixed: see Im et al. 1995c),
   we expect that about 4 -- 20 \% of the present day galaxies are
 produced via  major merging since $z=1$, adopting
 the present day  merger rate to be $0.005 \sim 0.023 Gyr^{-1}$ (Carlberg 1995)
 and the look-back time of $\sim 9 Gyr$.
  Today, roughly 20 -- 25 \% of galaxies in the volume limited sample
 are E/S0s (Buta et al. 1994; Lauberts \& Valentijn 1989), thus we expect
 the number evolution of more than 16 -- 100 \% to be observed
 if the ellipticals are the products of major merging.
   Similarly, we get the upper limit
 of the ``major merger rate'' (defined here the fraction of interacting pairs
 per unit time)  to be roughly $\lesssim~0.01 Gyr^{-1}$
 using our lower limit of $\phi_{*}(z=1) > 0.7 \phi_{*}(z=0)$.
  If the merger rate evolved as strongly as $\sim (1+z)^{4}$, the number
density
 of ellipticals at z=1 would be much less than 70 \% of the present day
 value. Hence models with the strong major merger rate evolution seem to be
 excluded (e.g., Carlberg 1992).

   However, the strong merger rate evolution with strong luminosity
 evolution could be reconciled with our data if the value of $\Omega$
 is locally low.
    The local low density universe has been suggested to
 explain the high value of $H_{0}$ with $\Omega_{m}=1$ (e.g., Wu et al.
 1995).  In order to reconcile the latest estimates of $H_{0}$ with the
$\Omega_{m}=1$ universe, one can assume that our local universe
($z < 0.2$) has a mass density only 10 $\sim$ 20 \% of the total mass
density of the universe.  Thus, at sufficiently high redshift ($z >
0.5$), the mass density needs to be about 5 to 10 times higher than that
in our local universe, and the same for the number density of galaxies
if the ratio of dark to luminous matter is roughly constant.  Our result
 appears to show the number density of E/S0s increases as a function
 of look-back time z=0 when $\Omega_{m}=1$,
 but not as much as the factor of 5--10.
    The local low density universe appears to be excluded,
 but the apparent deficit of E/S0s at $z \sim 1$ could simply be
 due to the strong number evolution caused by the strong major merger rate
 evolution. In that case,  the local low density universe $with$ the strong
 merger rate evolution may be consistent with our data
 (But for observational evidences against the strong major
 merger rate evolution, see Neuschaefer et al. 1995, 1996; Woods et al. 1995).

\section{Conclusions}

 We have constructed the luminosity functions of elliptical galaxies
using data from HST MDS and archived HST surveys.  Redshifts of these
E/S0s have been determined using $V-I$ colors and sizes to an accuracy
of $\sim 0.1$ up to $z \sim 1$.  From the constructed luminosity
functions, we find luminosity evolution of about $1 \pm 0.5$ mags at $z
\sim 1$.
  On the other hand, we exclude strong number evolution of
elliptical galaxies, and our data are consistent with a merger rate of
$\lesssim 0.01~Gyr^{-1}$ with very weak major merger rate unless
 the mass density of the universe is locally low.
  This latter result supports our earlier findings (Im et
al. 1995c) on the basis of the size and colour distributions of E/S0s.

\acknowledgements

This paper is based on observations with the NASA/ESA Hubble Space
Telescope, obtained at the Space Telescope Science Institute, which is
operated by the Association of Universities for Research in Astronomy,
Inc., under NASA contract NAS5-26555.  The HST Medium Deep Survey is
funded by STScI grants GO2684 {\it et seqq.}  We would like to thank the
other members of the Medium Deep Survey Team at JHU, especially Lyman
W. Neuschaefer and Eric J. Ostrander for their data reduction and
analysis work on the MDS pipeline, Richard Green for his help on collecting
 the redshift data, and Avi Naim for a careful reading of the manuscript.
  We are grateful to Simon Lilly for providing a machine
readable list of galaxies and redshifts from the Canada-France Redshift
Survey.

\clearpage

\begin{figure}[tb]
\centerline{\psfig{figure=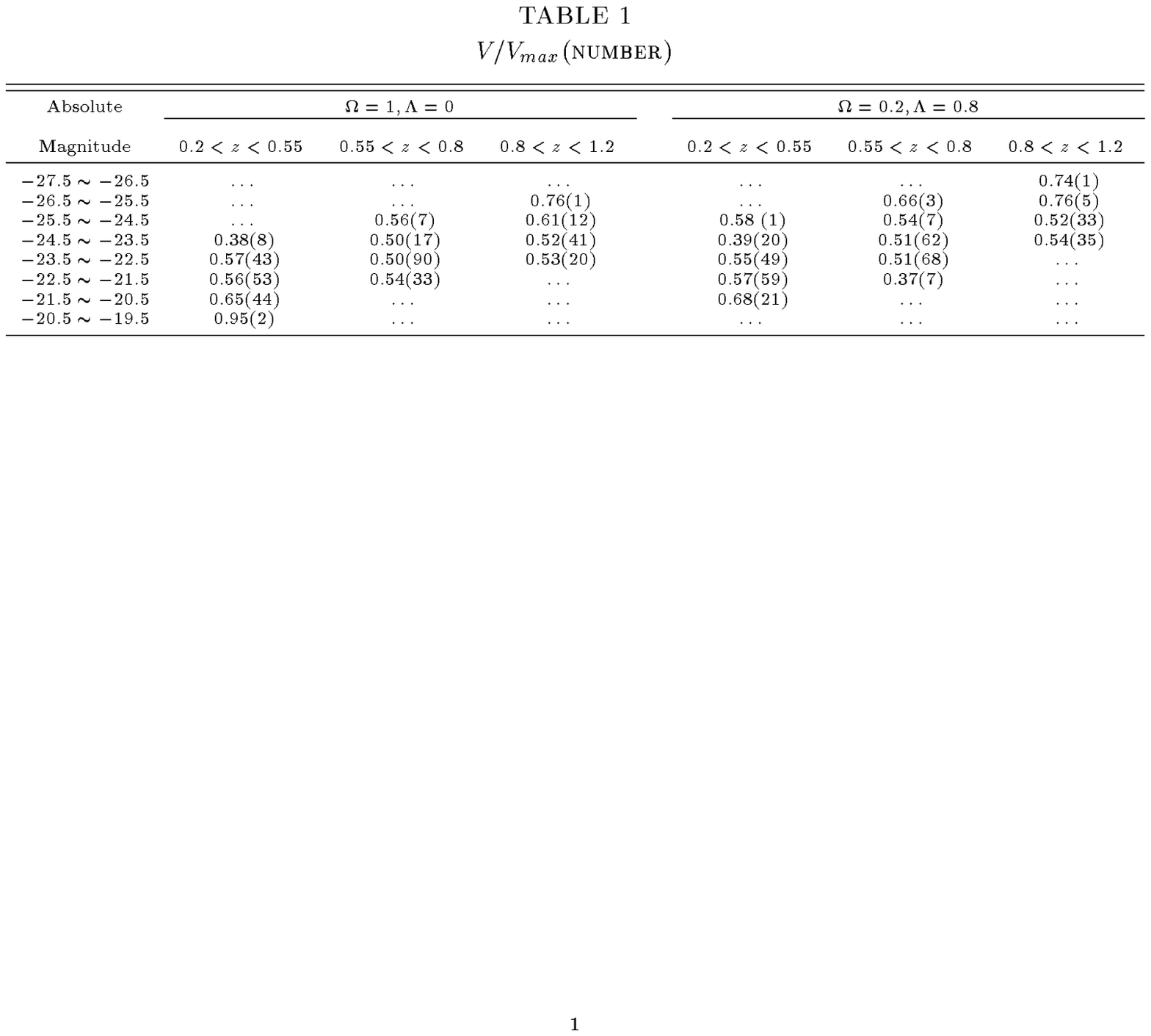}}
\end{figure}

\begin{figure}[tb]
\centerline{\psfig{figure=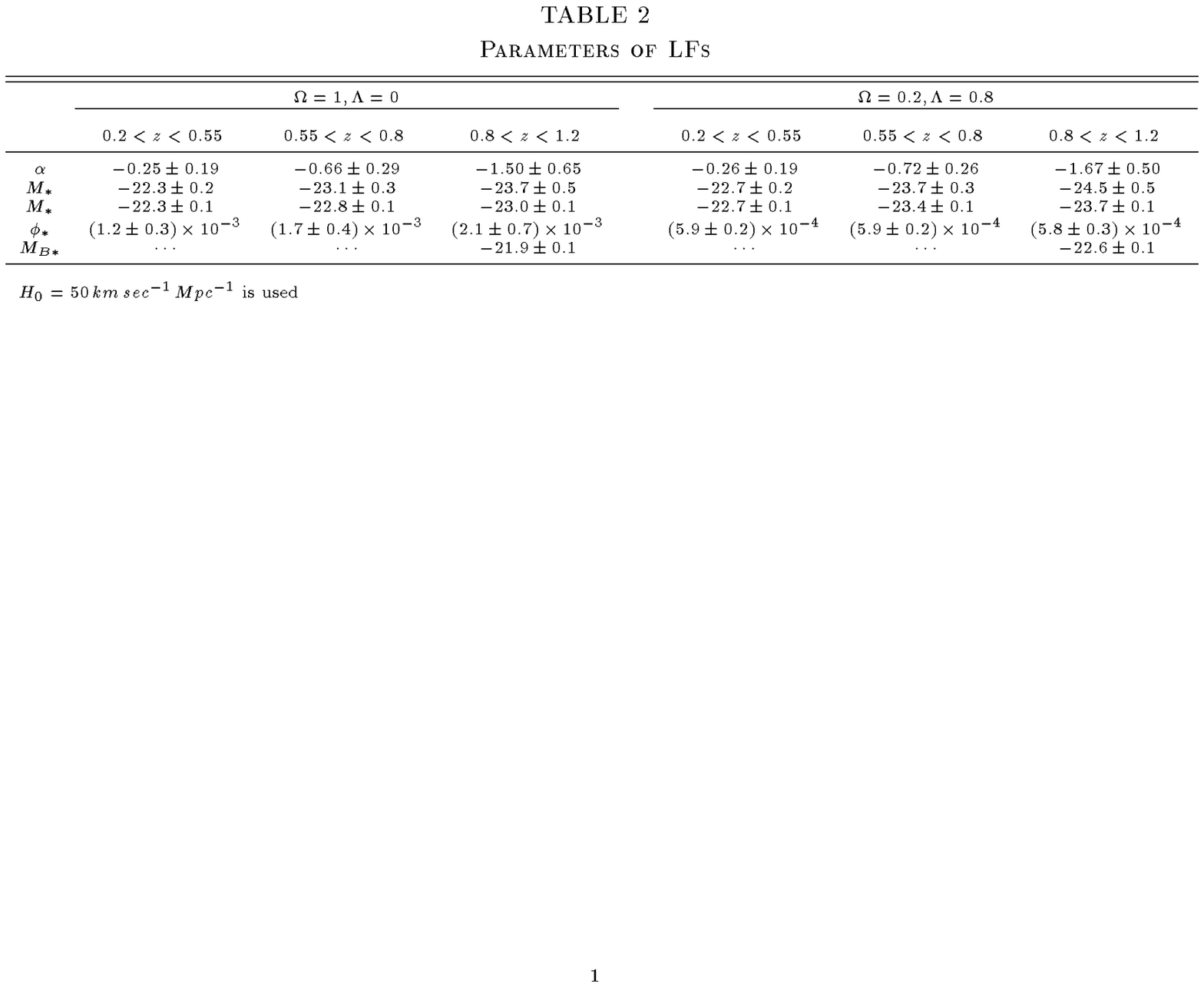}}
\end{figure}

\end{document}